# Real-Time Estimation of Equivalent Series Resistance for Predicting Output Capacitor Failures in Boost Converters

Antonino Pagano

Department of Engineering, University of Palermo, Palermo, Italy

**Abstract:** Output capacitors are among the most critical components in power electronic converters, as their degradation directly affects system stability and reliability. A key indicator of capacitor health is the Equivalent Series Resistance (ESR), whose progressive increase is strongly correlated with aging and imminent failure. This paper presents a real-time technique for estimating the ESR of output capacitors in boost converters, with the aim of enabling early fault prediction and condition-based maintenance. The proposed method leverages online parameter estimation to extract ESR values without interrupting converter operation. The estimation algorithm has been implemented on a low-cost STM32 Nucleo platform, demonstrating both computational efficiency and suitability for embedded applications. Experimental validation confirms that the approach provides accurate ESR tracking under varying load and operating conditions, allowing timely detection of abnormal capacitor behavior and preventing unexpected system failures.

## 1. Introduction

In recent times, increasing attention has been paid to self-diagnostic methods for DC-DC converters. Failures in electronic power converters are often due to the degradation of passive components, which can cause the entire system to malfunction and thus lead to a complete shutdown of the equipment. For this reason, it is very important to constantly evaluate passive components and monitor their 'health' by observing certain important parameters. A predictive self-diagnostic system improves system reliability as it can predict the failure of certain components and identify them; as a result, rapid replacement of components allows the system to resume operation quickly and avoid more serious problems.

The method used in this work to evaluate the correct functioning of DC-DC converters basically performs diagnostics by estimating circuit parameters. In particular, the health status of the electrolytic capacitor is monitored by estimating the ESR (*Equivalent Series Resistance*) [1], [2].

This experimental work builds upon a series of preliminary studies aimed at developing diagnostic techniques for power electronic converters. In particular, a boost converter was designed and a feasibility study was conducted, where simulations were employed to identify an appropriate diagnostic method for implementation on a microprocessor-based system. The selected approach estimates the circuit parameters using the least squares algorithm to minimize the error magnitude. The method was subsequently implemented on a microcontroller development platform, enabling practical validation of the proposed diagnostic technique.

## 2. Boost converter design

To test the self-diagnostic system, it was decided to design a dedicated boost converter with the following specifications:

- Boost converter input voltage $V_{in} = 12\ V$;
- Boost converter output voltage $V_{out} = 20\ V$;
- Switching frequency $f_{sw} = 10\ kHz$;
- Duty cycle $D = 0.4$;
- Load resistance $R = 20\ \Omega$;
- Output voltage ripple $\Delta V_{out} = 0.25\ V_{pp}$ ;
- Maximum current ripple on the inductance $\Delta I_L = 2\ A$;

The deterioration of the capacitor, with a consequent increase in ESR, was reproduced by adding a network of five $200\ m\Omega$ resistors that could be connected in series to the capacitor via jumpers. *Jumpers* could also be used to vary the filter capacity value in order to verify the accuracy of the estimate under various conditions. The circuit constructed is shown in Fig. 1.

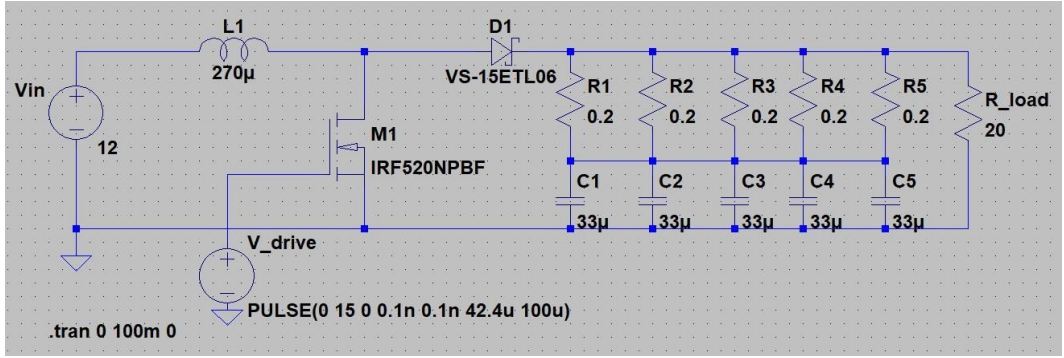
*Figure 1: Circuit diagram of the boost converter implemented on LTspice.*

## 3. Algorithm for estimating circuit parameters in DC-DC converters

The diagnostic technique chosen is *signal processing*, which uses the least squares method to reduce the error in the estimation of values [3].

The main objective of the algorithm is to estimate the following quantities: load resistance $R_{load}$, inductor value $L$, output capacitance $C$ and its parasitic resistance $ESR$. This technique is very accurate, but to maintain its precision, the sampling frequency of the acquired signals must be in the MHz range. The steps of the algorithm are as follows:
- The necessary signals ($I_L$, $V_o$, $V_C$, $V_{MOS}$) are sampled during a single switching period.
- The collected samples are grouped as data vectors of the algorithm. The average values are calculated.
- An analysis of the vectors is performed in order to identify the operating states of the circuit, distinguishing between the data sampled during $T_{on}$ and during $T_{off}$. - An evaluation of the load resistance is performed.
- The relationship linking the input variables to the ESR value is applied.
- Considering the samples of the vectors $I_L$ and $V_{out}$ during $T_{on}$, the inductance and capacitance are evaluated.

By way of example, the formulas used to calculate the load resistance and ESR are shown below.

$$R_{load} = \frac{V_{media}}{I_{L\,media}(1-D_{on})} \qquad ESR = -\frac{V_{out}\left(\frac{T_{on}}{2}\right) - V_{media}}{I_{L\,media}}$$

## 4. Hardware description

Fig. 2 shows a simplified block diagram of the entire system. The microcontroller acquires four signals from the boost converter, executes the algorithm for estimating the circuit parameters and sends the data obtained to a notebook via the serial connection. The system also generates a PWM signal to control the converter.

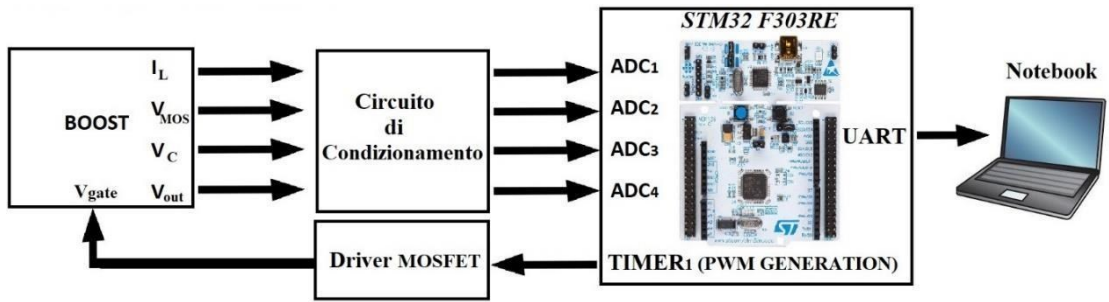
*Figure 2: Block diagram of the system implemented.*

The four signals acquired are the current on the inductor ($I_L$), the voltage on the MOSFET ($V_{MOS}$), the voltage on the capacitor ($V_C$) and the output voltage ($V_{out}$). Fig. 3 shows the system implemented; the printed circuit board was created using the *Press-N-Peel* technique.

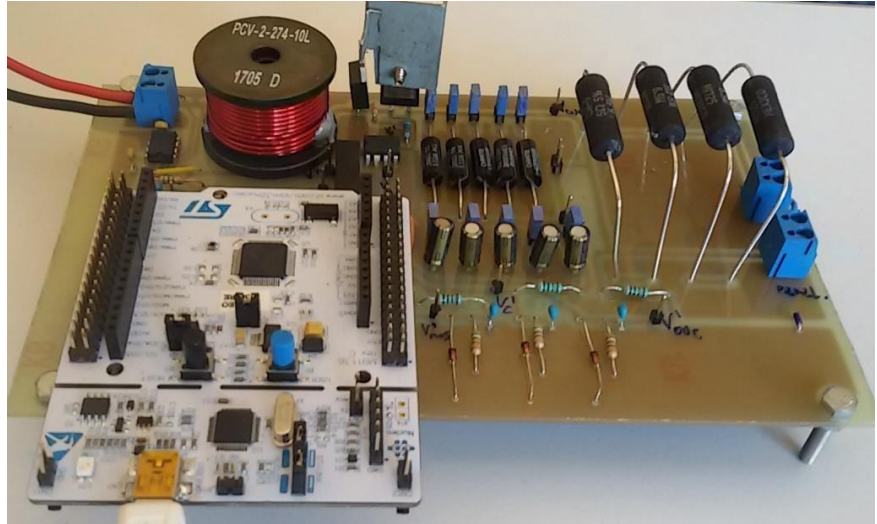
*Figure 3: Photo of the system implemented.*

**5. Results obtained** Tables 1 and 2 show the results of the algorithm as the ESR varies for two different capacitance values. The estimates of the circuit parameters shown in the table were obtained by calculating the average value of 20 successive acquisitions. The results also show the variance.

| ESR-Value entered ($m\Omega$) | ESR-Estimated value ($m\Omega$) | Estimate of $R_{load}$ ($\Omega$) | Estimate of $C$ ($\mu F$) | Estimate of $L$ ($\mu H$) |
|---|---|---|---|---|
| 40 | 40 ± 9 | 20.031 ± 0.015 | 103 ± 5 | 276 ± 2 |
| 50 | 55 ± 9 | 20.022 ± 0.021 | 109 ± 4 | 267 ± 3 |
| 70 | 78 ± 8 | 20.035 ± 0.018 | 112 ± 6 | 266 ± 3 |
| 100 | 106 ± 10 | 20.026 ± 0.012 | 108 ± 6 | 267 ± 2 |
| 200 | 202 ± 10 | 20.075 ± 0.006 | 109 ± 4 | 261 ± 3 |

*Table 1: Estimates of circuit parameters as ESR varies for C = 99µF.*

| ESR-Value entered ($m\Omega$) | ESR-Estimated value ($m\Omega$) | Estimate of $R_{load}$ ($\Omega$) | Estimate of $C$ ($\mu F$) | Estimation of $L$ ($\mu H$) |
|---|---|---|---|---|
| 40 | 44 ± 9 | 20.13 ± 0.05 | 161 ±12 | 269 ± 2 |
| 50 | 55 ± 8 | 20.10 ± 0.02 | 156 ± 10 | 268 ± 3 |
| 70 | 79 ± 10 | 20.11 ± 0.02 | 159 ± 12 | 266 ± 5 |
| 100 | 107 ± 9 | 20.09 ± 0.02 | 155 ± 11 | 267 ± 2 |
| 200 | 209 ± 11 | 20.13 ± 0.02 | 155 ± 10 | 265 ± 5 |

*Table 2: Estimates of circuit parameters as ESR varies for C = 165µF.*

To obtain the ESR estimates shown in the table, a constant offset was subtracted from all calculated values in order to obtain values similar to the value of the resistors connected in parallel; this offset is linked to the ESR of the capacitor and the parasitic resistances of the tracks and *jumpers*. It can be seen that the implemented algorithm is able to follow the variations in ESR and, consequently, provides information on the degradation of the capacitance. As proof of this, Fig. 4 shows the linear regression lines for the estimates made on the ESR. The linear regression lines shown in Fig. 4 were obtained for 5 different capacitance values; the graph also shows an increase in the initial offset (not subtracted here) in the estimated ESR value as the capacitance decreases capacity.

This phenomenon can be explained by the fact that the variation in capacity is achieved by inserting capacitors in parallel. When the 5 capacitors are all inserted, the contribution of the parasitic resistances of the circuit (tracks, solder joints, jumpers) is reduced, as these components are connected in parallel.

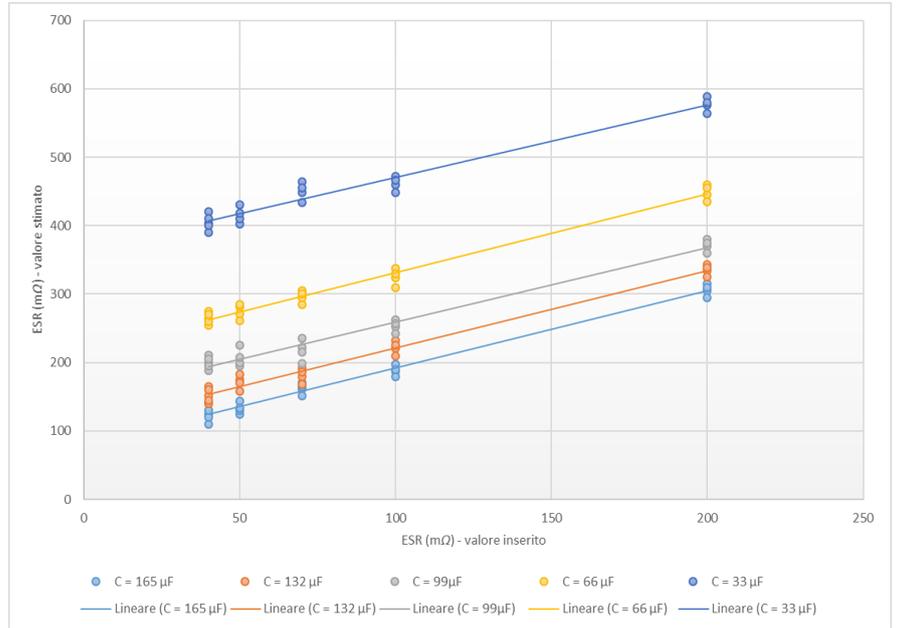

*Figure 4: Linear regression lines on ESR estimates.*

Capacitor deterioration not only causes an increase in ESR, but also a simultaneous reduction in capacitance. Table 3 and Fig. 5 show how the algorithm is also able to track capacitance variations, thus demonstrating its ability to indicate capacitor deterioration.

| Entered value (µF) | Estimated value (µF) | Variance |
|---|---|---|
| 165 | 175 | ±13 |
| 132 | 141 | ±10 |
| 99 | 104 | ±5 |
| 66 | 67 | ±3 |
| 33 | 35 | ±2 |

*Table 3: Capacity estimates detected by the algorithm.*

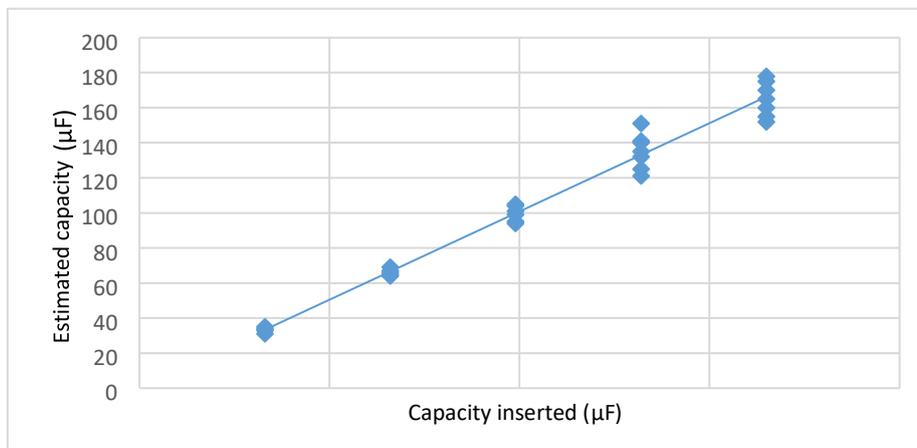

*Figure 5: Linear regression line on capacity estimate.*

Both Table 3 and Fig. 5 show a reduction in variance as the capacitance value decreases; that is, the lower the capacitance, the closer the estimate is to the true value. This observation can be justified by the fact that for large capacitance values, the voltage *ripple* amplitude is lower and, consequently, the capacitance estimate, which is based on *the ripple* measurement, is affected by the presence of noise. For low capacitance values, on the other hand, the *ripple* amplitude is greater and the estimate is more reliable. Several methods can be used to make the estimate of the value of C more robust. For example, A-D converters with higher resolution (24 bits) can be used and then digital filtering can be applied to remove noise. Another solution could be to separate the AC component from the DC part and then amplify the AC part to make it compatible with the dynamics of the analogue-to-digital converter used.

## 6. Conclusions

In this work, a boost converter and a diagnostic system combined with it were created, implemented on the *Nucleo STM32F303RE* development board, capable of controlling any variations in the circuit parameters of the converter due to the degradation of passive components. The algorithm adopted is based on the least squares method and is capable of estimating circuit parameters, including the value of the equivalent series resistance of the capacitance and the value of the capacitance itself [4]. This diagnostic technique is low cost; in fact, the algorithm can be implemented in the same microprocessor system dedicated to controlling the converter, thus eliminating the need for additional sensors. The technique has been implemented and verified experimentally, with satisfactory results for the estimated parameters. Although the technique has been applied to a boost converter, the algorithm can be easily modified for other converter topologies.